\documentclass{aastex61}

\received{}
\revised{}
\accepted{}
\submitjournal{ApJ}

\shorttitle{O$_2$ in Mrk 231}
\shortauthors{Junzhi Wang et al.}

\begin{document}

\title{Molecular Oxygen in the nearest QSO Mrk 231}

\correspondingauthor{Junzhi Wang}
\email{jzwang@shao.ac.cn}
\author
{Junzhi~Wang}
\affiliation{Shanghai Astronomical Observatory, Chinese Academy of Sciences,80 Nandan Road, Shanghai, 200030, China}
\affiliation{Key Laboratory of Radio Astronomy, Chinese Academy of Sciences,   Nanjing, 210008,  China}

\author{Di Li}
\affiliation{CAS Key Laboratory of FAST, National Astronomical Observatories, Chinese Academy of Sciences, Beijing 100012, China}
\affiliation{University of Chinese Academy of Sciences, Beijing 100049, China}

\author{Paul F. Goldsmith}
\affiliation{Jet Propulsion Laboratory, California Institute of Technology, 4800 Oak Grove Drive, Pasadena, CA 91109, USA}

\author{Zhi-Yu Zhang}
\affiliation{Institute for Astronomy, University of Edinburgh, Royal Observatory,  Blackford Hill, Edinburgh EH9 3HJ, UK}
\affiliation{ESO, Karl Schwarzschild Strasse 2, D-85748 Garching, Munich, Germany}
\affiliation{School  of Astronomy and Space Science, Nanjing University, Nanjing,  210093, China}

\author{Yu Gao}
\affiliation{Department of Astronomy, Xiamen University, Xiamen, Fujian 361005, China}
\affiliation{Purple Mountain Observatory, Chinese Academy of Sciences, 2 West Beijing Road, Nanjing, 210008, China}
\affiliation{Key Laboratory of Radio Astronomy, Chinese Academy of Sciences,   Nanjing, 210008,  China}

\author{Yong Shi}
\affiliation{School  of Astronomy and Space Science, Nanjing University, Nanjing,  210093, China}

\author{Shanghuo Li}
\affiliation{Shanghai Astronomical Observatory, Chinese Academy of Sciences,80 Nandan Road, Shanghai, 200030, China}
\affiliation{Key Laboratory of Radio Astronomy, Chinese Academy of Sciences,   Nanjing, 210008,  China}

\author{ Min Fang}
\affiliation{Department of Astronomy, University of Arizona,  933 North Cherry Avenue, Tucson, AZ 85721, USA}

\author{Juan Li}
\affiliation{Shanghai Astronomical Observatory, Chinese Academy of Sciences,80 Nandan Road, Shanghai, 200030, China}
\affiliation{Key Laboratory of Radio Astronomy, Chinese Academy of Sciences,   Nanjing, 210008,  China}

\author{Jiangshui Zhang}
\affiliation{Center For Astrophysics, GuangZhou University, GuangZhou  510006, China}

\begin{abstract}

We report the detection of  an emission feature at the 12$\sigma$ level with FWHM line width of about 450 km s$^{-1}$  toward the nearest quasi-stellar object, QSO Mrk 231.   Based on observations with the IRAM 30 m telescope and the NOEMA Interferometer,  the  $1_1$-$1_0$  transition of molecular oxygen is the likely origin of  line   with rest frequency close to 118.75 GHz.  The velocity of the O$_2$ emission in Mrk 231 coincides with the red wing seen in CO emission, suggesting that it is associated with the outflowing molecular gas, located mainly at  about ten kpc away from the central AGN. This first detection of extragalactic molecular oxygen provides an ideal tool to study AGN-driven molecular outflows on dynamic time scales of tens of Myr. O$_2$ may be a significant coolant for  molecular gas in such regions affected by  AGN-driven outflows. New astrochemical models are needed to explain the implied high molecular oxygen abundance in such regions  several kpc away from the center of galaxies.

\end{abstract}

\keywords{ISM: abundances; quasars: individual: Mrk 231}

\section{Introduction} \label{sec:intro}

As the third most abundant element in the universe after hydrogen and helium, oxygen and its chemistry in dense interstellar  clouds  are important for understanding the properties of molecular gas  \citep{Goldsmith2011,Hollenbach2009}.   Due to the attenuation of earth's atmosphere, it is impossible to observe  O$_2$ lines near  their rest frequencies from the ground. Searches for O$_2$ emission in  the Milky Way  have been carried out from orbital observatories including the Submillimeter Wave Astronomy Satellite (SWAS)  \citep{Goldsmith2000, Goldsmith2002}, Odin \citep{Liseau2005, Larsson2007}, and {\it Herschel} \citep{Goldsmith2011, Liseau2012}.  Based on  observations with those space missions (SWAS, Odin, and  {\it Herschel}), the [O$_2$]/[H$_2$] abundance ratios in dense gas regions  are more than two orders of magnitude lower than predictions  from pure gas-phase chemical models \citep{Langer1989, Bergin1998},   $\sim10^{-5}$ to $\sim10^{-4}$.  Because of its low abundance and weak line emission, O$_2$ is  believed not to be an important coolant of the gas in dense  molecular clouds.  

The O$_2$ line in extragalactic sources is red-shifted away from the  attenuation of the Earth's atmosphere, and thus can be observed with ground-based millimeter facilities. However,  none of the observations detected O$_2$ emission in the  galaxies observed  \citep{Liszt1985,Goldsmith1989, Combes1991, Frayer1998, Kanekar2015} in the past two decades, with the best  [O$_2$]/[H$_2$]  upper limit  being 1$\times$10$^{-6}$  in NGC 6240 at the 1$\sigma$ level  \citep{Combes1991}, while the non-detection of absorption features toward a foreground galaxy in front of the BL Lac object  B0218+357  give an upper limit  of  2$\times$10$^{-7}$  at the 1$\sigma$ level  \citep{Combes1997}.  Although the freeze-out of oxygen carriers, particularly H$_2$O, onto dust grains is generally cited as the explanation for the lack of  gas-phase O$_2$, a comprehensive picture of oxygen chemistry  in different interstellar environments is still missing.

 The enhancement of O$_2$ emission in Orion \citep{Goldsmith2011, Chen2014} is best explained by recent passage of a shock, while  the AGN-driven molecular outflow can produce continuous shocks in associated molecular clouds. Thus,  Mrk 231 is a good choice  to  search for extragalactic  O$_2$ emission.
At a distance of 172 Mpc  \citep{Fischer2010}, Mrk 231 is the nearest QSO and the most luminous ULIRG (Ultra-Luminous InfraRed Galaxy, $L_{IR}>10^{12}L_\odot)$ in the local universe \citep{Feruglio2010}.  It contains high velocity molecular outflows revealed by far infrared OH absorption \citep{Fischer2010}, and  millimeter emission of  CO \citep{Feruglio2010, Cicone2012} and  is observed in dense gas tracers  \citep{Aalto2012}.  The  molecular  mass outflow rate  in Mrk 231 is estimated to be 700 $M_\odot$/year based on CO J=1-0 observations, which is higher than the star formation rate of $\sim$ 200  $M_\odot$/yr \citep{Feruglio2010}.

The O$_2$ $N_J$=$1_1$-$1_0$ transition with  rest frequency  118.750343 GHz, is free from contamination by known nearby lines \citep{Sandqvist2008}.  This spectral line, which can  be observed toward local galaxies  with red shift greater than $\sim$0.025  using  ground based millimeter-wavelength facilities, is the best choice for a search for extragalactic molecular oxygen among  O$_2$ transitions.  It was also the choice for the previous searches for extra-galactic O$_2$ emission referred to above.  

In this paper, we describe the observations and data reduction in \S2, present the main results  in \S3,  discuss the results in \S4, and give as brief summary and discuss future prospects in \S5.

\section{Observations and data reduction}

\subsection{Observation with IRAM 30 meter telescope and data reduction }

The observations were carried out from August 20 to 23,  2015, with the IRAM 30m telescope in good  weather conditions  with precipitable water vapor (pwv) less than 4mm (Project id: 068-15, PI: Junzhi Wang).  The Eight MIxer Receiver (EMIR) with dual polarizations, Fast Fourier Transform Spectrometer (FTS) backend, and  standard wobbler switching mode at 0.5Hz with $\pm120''$ offset beam throw, were used toward Mrk 231  pointing at RA=12:56:14.2  DEC=+56:52:25.0 (J2000). The beam size of the IRAM 30m telescope at the observing frequency ($\sim$114 GHz) is about 23$''$.    In order to verify that the signal is from  the sky  rather than  being radio-frequency interference (RFI) at the IF  frequency or from the backend, four  local oscillator (LO) tuning  setups were used during the observations, which are 103.77 GHz, 103.87 GHz, 103.57 GHz, and 103.07 GHz for August 20, 21, 22, 23, respectively.  We dumped the data every 1.7 minutes as one scan,  and did  one calibration observation every 6 scans.  The  CLASS package GILDAS \footnote{http://www.iram.fr/IRAMFR/GILDAS} was used for data reduction. During the data reduction,  we checked the spectrum of each scan and dropped  scans with  bad baseline and/or hot spots in some channels, which comprised  about 5\% of the total data.  We combined the data from  all four days to obtain the final spectrum  weighted by the  rms of each individual scan. The rms of the final spectrum is 0.38 mK in T$_{mb}$ after smoothing to the velocity resolution of 98.7 km s$^{-1}$, while the total effective observing time (on+off) is 408 minutes.

 \subsection{Observations with NOEMA and data reduction}

The NOEMA observations were carried out  on December 1 and 3, 2017, (Project id: d17ae001, PI: Junzhi Wang)  under good weather conditions with the same pointing center as that of IRAM 30m observation in 2015.  The frequency coverage of upper side band (USB) is from about 108.2 GHz to 116.2 GHz, which covers  the redshifted CN 1-0, CO 1-0,  O$_2$, and  HC$_3$N $J$=13-12  lines.  The original frequency resolution was 2 MHz, which corresponds to 5.26 km s$^{-1}$ at 114 GHz.   On Dec 1,   3C84 was used as flux and bandpass calibrator, while   1300+580 and   1150+497 were used as  phase calibrators, with  7 antennae in the array  and  a total of 2.6 hours time on Mrk 231.  There were two observing blocks on Dec 3 using 3C273 as flux and bandpass calibrator while the phase calibrators were again  1300+580 and   1150+497. One  block was done with 8 antennae in the array  and  a total of 0.8 hours  on Mrk 231, while the other was done with 9 antennae in the array  and  a total  of 3.8 hours on Mrk 231. 

 After accounting for bandpass, flux, phase and amplitude calibrations  for the visibility  data with `CLIC' in GILDAS,  we carried out imaging and deconvolving with both `MAPPING' in GILDAS and also CASA. The final results from both software packages are consistent.  The images and spectra presented in this paper were  done with CASA after smoothing to 10 MHz frequency resolution corresponding to a velocity resolution of 26.3 km s$^{-1}$, and giving a slightly noncircular beam  of  3.63$''\times3.32''$ with $PA$=$-65^{\circ}$.   
 
\section{Results}

\subsection{Detection of  O$_2$ $N_J$=$1_1$-$1_0$ emission with the IRAM 30 meter telescope} 

An emission feature was detected at about  the 12$\sigma$ level, which we identify as  O$_2$ $N_J$=$1_1$-$1_0$  emission with similar velocity range  to that of  the red wing of  outflowing gas observed in CO $J$=1-0.  The velocity range of this feature is about 150 to  650 km s$^{-1}$  (see Figure~1).
The  frequency resolution of the spectrum   shown  in Figure 1 is   37.5 MHz, which  corresponds to 98.7 km s$^{-1}$ at 113.945 GHz for  the O$_2$ $N_J$=$1_1$-$1_0$ line with $z$=0.04217 \citep{Bryant1996,Carilli98}.  This red shift corresponds to 12642 km  s$^{-1}$ as the optically-defined velocity   for this galaxy from NASA/IPAC Extragalactic Database (NED, http://ned.ipac.caltech.edu).

The velocity integrated flux of O$_2$ $N_J$=$1_1$-$1_0$ emission is 0.88$\pm0.07$ K km s$^{-1}$, which corresponds to  4.4$\pm0.35$ Jy km s$^{-1}$ with the conversion factor of 5.0 from K to Jy. The error of 0.07 K km s$^{-1}$  is from  single component Gaussian fitting for the spectrum, while there will be about 20\% more uncertainties due to absolute flux calibration which should be considered if comparing with different observations.  The flux of  detected  O$_2$ emission  in Mrk 231 is comparable to that of the red wing of CO 1-0 previously detected \citep{Cicone2012}, which  is   2.32$\pm0.17$ Jy km s$^{-1}$. Our CO $J$=1-0 data  obtained simultaneously with the O$_2$ in our IRAM 30 m observations provided similar results for the red wing of CO 1-0, with similar flux level and velocity range (see Figure~1).   In addition to O$_2$ $N_J$=$1_1$-$1_0$ line, we also detect   HC$_3$N $J$=13-12 line,  at a significance of 8 $\sigma$. This line is at $\sim$1200 km s$^{-1}$ in the Figures since the O$_2$ $N_J$=$1_1$-$1_0$ frequency corrected central velocity of the galaxy  was used as reference frequency. The data are shown at higher velocity resolution and wider velocity coverage in Figure 2.

Since the on-source times on Aug 20 and  21 are about half of those on Aug 22 and Aug 23, and the system temperatures on Aug 20 and 21 are  higher than those  on Aug 22 and 23, most of which are lower than 300 K, the main contributions to the final spectrum are from the data on Aug 22 and 23. We present the  data of Aug 22 and 23 as the two spectra in Figure 3. Even though the noise levels of the two spectra  are much higher than the 4-day combined spectrum, the two  emission features, especially the one in the range $\sim$ 100 to 800 km s$^{-1}$, can be seen in the individual spectra shown in  Figure 3.  On the other hand, due to very little on-source time in Aug 20 and 21 without any emission signature, the spectra are not shown here.

\subsection{Emission features confirmed with the NOEMA interferometer}

The single dish detections of O$_2$  $N_J$=$1_1$-$1_0$  and HC$_3$N $J$=13-12 emission in Mrk 231 were confirmed with  observations using the NOrthern Extended Millimeter Array (NOEMA).  O$_2$ $N_J$=$1_1$-$1_0$ emission detected with the NOEMA extends to $\sim15''$  (1$''$=0.87kpc  for Mrk 231) from the center of galaxy (see Figure 4). After the inclination correction with the average value of cos($\theta$) for random sample, the  physical projection will be 15$\times0.87\times\frac{\pi}{2}=20.5$kpc. However, since the real inclination is unknown, the distances to center in the discussion part are without such correction. The total flux of the identified emission features observed with NOEMA  (see Table 1)  is 1.9$\pm0.2$  Jy km s$^{-1}$, which includes about 43\% of the emission detected with the 30 m telescope,  and covers a  consistent velocity range,  from about 180 to 570 km s$^{-1}$.    Since the signal is not a point source located at the phase center, which leads to the difficulty of obtaining the emission features in UV plane data, UV spectrum was not presented.

Continuum subtraction was implemented in the cleaned image data cube based on nearby  line free channels, from about -2500 to -1000 km s$^{-1}$ and from about 1600 to 3000 km s$^{-1}$. HC$_3$N 13-12  was seen  as compact emission in the nuclear region, coincident with the AGN/extreme starburst and consistent with the spectrum taken by the IRAM 30 m telescope, while  O$_2$ emission was observed to be extended in outer disk regions $\sim$10 kpc from the nucleus.  The emission was detected in several different regions with different velocities (see Figures 4 and 5). The flux of the HC$_3$N 13-12 line detected with the NOEMA  is 1.55$\pm$0.07 Jy km s$^{-1}$, which agrees with the total emission detected with the IRAM 30 m single-dish telescope within the uncertainties of the observations at the two facilities.

 We present the velocity-integrated maps for different velocity ranges  in Figure 5, smoothed to about 150 km s$^{-1}$ velocity resolution.  The noise level (1$\sigma$)  is about 0.045 Jy km s$^{-1}$  for each map, while the contours are from 2$\sigma$   with 1$\sigma$ steps. Emission features can be seen in several velocity components above 3$\sigma$, especially from 180 to 570 km  s$^{-1}$, which is consistent with the emission feature detected with the IRAM 30 m telescope.  The O$_2$ moment zero map (Figure 4, in both gray scale and contours) includes emission features in the velocity range 180 to 570 km s$^{-1}$. The corresponding angular regions are marked in Figure 5.   With information from  NOEMA observations, even the velocity range of O$_2$ $N_J$=$1_1$-$1_0$ emission is similar to that of CO red wing, which traces  molecular outflows within 3$''$ \citep{Feruglio2010, Cicone2012},  O$_2$ is from different regions of outflowing gas  traced by CO red wing.

\subsection{Line identification as O$_2$ $N_J$=$1_1$-$1_0$ emission}

 The image rejection of the EMIR receiver on the  IRAM 30 meter telescope is better than 13 dB, which corresponds to $\ge$ 20:1.  
  The lower sideband image frequency corresponding to the feature of interest in the upper sideband is different on different days, since different LO  frequencies were used. 
 The result is that any signal from the lower sideband will move around in frequency rather than adding together when the data from different days are added.
 The image frequencies are in the range 96 GHz to 97.5 GHz during the four days, and are not near the strong dense gas tracer CS $J$=2-1 at the rest frequency of 97.980953GHz. Even the strongest dense gas tracer in the 3mm band, HCN 1-0,  produced an antenna temperature of  only 6.5 mK in Mrk 231 with the  IRAM 30 meter telescope \citep{Jiang2011}.   
The  lower sideband image frequencies with NOEMA corresponding to the upper sideband emission feature identified as O$_2$ $N_J$=$1_1$-$1_0$ are  between 92 to 94 GHz (due to the different IF frequency), and there were no emission features detected in the lower sideband, while the detected CS $J$=2-1 line with NOEMA observation is about half the intensity of HCN $J$=1-0.   
 The expected confusion from the lines in the image sideband should thus be much less than 0.2 mK for the final spectrum, which does not  affect the  line identification. 
 

The rest frequency range of the  emission feature from -100 to 900 km s$^{-1}$ in Figure 1 is  from about 118.4 GHz to 118.8 GHz.  From  splatalogue (http://www.cv.nrao.edu/php/splat/), which combines  the JPL molecular  catalog \citep{Pickett1998}  and  the CDMS catalog \citep{Muller2001},  there are 350 lines in this frequency range, including  the O$_2$ $N_J$=$1_1$-$1_0$ line at 118.75034 GHz, which is the only one that  has been detected in space \citep{Larsson2007}. With the criterion of lower level energy less than 100 K, the number of lines is decreased  to 83, which are mainly from  complex molecules, such as CH$_3$OCH$_3$ and CCCN, or  isotopologues, such as $^{33}$SO$_2$. The abundances of such complex molecules and isotopologues are not be expected to be high enough to produce emission stronger than HC$_3$N $J$=13-12 in Mrk 231. Otherwise, such features should be easily detected in Galactic molecular clouds, while  HC$_3$N $J$=13-12 was the strongest (and possibly the only) emission line detected in the frequency range from 117.8 GHz to 118.75 GHz toward the Sgr A complex region in the Galactic centre  with Odin \citep{Sandqvist2008}. 

 If we restrict the emission feature  to fall within $\pm$100 km s$^{-1}$ of the central velocity of the galaxy, with the rest frequency range  from 118.54 to 118.62 GHz, there are 57  lines from splatalogue, none of which has been reported as detected. All of them, except for SO$_2$ at 118.57743 GHz and the vibrationally excited HC$_3$N $J$=13-12 1e line at 118.56145 GHz,  are extremely  complex molecules with more than 5 atoms. The lower level energy of SO$_2$ and HC$_3$N transitions are from about 250K to 1600K, which makes them unlikely candidates for the observed emission. In addition, the vibrationally excited line would almost certainly be significantly weaker than the corresponding ground vibrational state line, as found even in regions of high infrared flux that can radiatively excite the vibrational levels e.g. \citet{Goldsmith82}.

 If the origin of  the emission were a low-lying transition of a complex molecule, it should have been detected in sources in the Milky Way. However, no such emission feature was reported by \cite{Sandqvist2008}.  Thus, the most probable   identification  of the emission feature from 100 to 800 km s$^{-1}$ is  O$_2$ $N_J$=$1_1$-$1_0$ with the  velocity of the red wing of the gas in Mrk 231.

\section{Discussion}

\subsection{O$_2$ abundance in different regions of  Mrk 231: outer disk v.s. central 2 kpc}

Stellar components  traced at optical wavelengths  can be seen \citep{2011ApJ...729L..27R} in the region with high velocity red-shifted O$_2$ emission, while studies of  molecular gas traced by CO lines    \citep{Feruglio2010, Cicone2012, Feruglio2015}  were  focused on the central 2 to 3 kpc region, even though some outflow features can be seen  up to 5$''$ ($\sim$4 kpc) north of the central region   \citep{Cicone2012}.
In the region with O$_2$ detection extending from 450 to 570 km s$^{-1}$,   the flux of  the simultaneously-obtained CO $J$=1-0 line was about 4 times that of the O$_2$  $N_J$=$1_1$-$1_0$ line. However,  the velocity range  of  CO 1-0   is primarily associated with the  central region  of Mrk 231, which is quite distinct   from the O$_2$ emission. 
Thus, even though  the O$_2$ emission has a similar velocity range as the red wing of molecular outflow traced by CO 1-0, it is coming from a different portion of Mrk 231.    
Molecular line emission of CH$^+$ $J$=1-0  without associated CO emission has been found in several high redshift starburst galaxies \citep{2017Natur.548..430F}, which indicates that  CO emission is not necessarily detectable  in the regions in which some special  molecules, such as O$_2$ and  CH$^+$ can be detected. 

The CO $J$=1-0  fluxes in the regions with  O$_2$ detections  were comparable to, or even less than that of O$_2$  $N_J$=$1_1$-$1_0$ line, if we  consider that  only CO 1-0 emission with the same velocity range as the   O$_2$ line  is from the same gas as responsible  for the  O$_2$  emission.   Using the same CO 1-0 flux to H$_2$ conversion factor in the O$_2$ emission region  as that of the host galaxy and   assuming  that the O$_2$ line is optically thin  with a 15 K excitation  temperature  ($T_{ex}$), which is the same as  was assumed in NGC 6240 \citep{Combes1991}, the   O$_2$ to H$_2$ abundance ratio  can be estimated to be higher than  $10^{-4}$  (see Table 2), nearly 2 orders of magnitude  higher than that of Galactic sources detected in O$_2$, and higher than  the CO to H$_2$ abundance ratio.  
The dependence of  the O$_2$ column density on its excitation temperature in LTE is $N(O_2)$ $\propto$ $T_{ex}^{0.67}$ \citep{Liseau2005}.
Thus if the excitation temperature of  O$_2$ were higher than 15 K, the O$_2$ to H$_2$ abundance ratio would be even higher.

The nondetection of  O$_2$ emission in the central 2 kpc of  Mrk 231 with our NOEMA observations gave the best upper limit at the of extragalactic  O$_2$ $N_J$=$1_1$-$1_0$  to CO 1-0  line ratio of 1$\times$10$^{-3}$ (1$\sigma$) , much lower than the 1.2$\times$10$^{-2}$ value for NGC 6240, which was used previously for estimating  the O$_2$/H$_2$  abundance ratio \citep{Combes1991}.  
Using the same  CO $J$=1-0 flux to H$_2$ conversion factor as that in NGC 6240 and similar excitation conditions of the O$_2$ molecules \citep{Combes1991}, the O$_2$ to H$_2$ abundance ratio should be less than 8$\times$10$^{-8}$ in the host galaxy of Mrk 231,  which is consistent  with that found the vast majority of that in dense  molecular gas in the Milky Way \citep{Goldsmith2000,Pagani2003}.  The only exceptions are Orion \citep{Goldsmith2011} (having significantly higher abundance in 10$^{-6}$ in a very small region) and $\rho$ Oph    \citep{Larsson2007,Liseau2012} (with comparable abundance 5$\times10^{-8}$).

We compiled the sources with  O$_2$ detections,  the best upper limit of  O$_2$ to H$_2$ abundance ratio in NGC 6240 with emission      \citep{Combes1991} and in B0218+357 with absorption \citep{Combes1997}   in the  literature,  and show the results in Table 2. 
If we wish to detect O$_2$ $N_J$=$1_1$-$1_0$ emission in extragalactic sources without O$_2$ abundance enhancement, observations of the red-shifted O$_2$ $N_J$=$1_1$-$1_0$ line  with 5 times  lower noise level than that of our observations with NOEMA toward Mrk 231 are needed. 
With the Atacama Large millimeter/submillimeter array  (ALMA), about ten hours on-source time toward nearby  gas-rich galaxies with bright CO emission  can achieve such sensitivity.

\subsection{Origin of molecular Oxygen: large scale shock caused by AGN molecular outflow?}

The velocity range of our  O$_2$ emission is mainly from red-shifted gas, especially from $\sim$345  to 615 km s$^{-1}$, with projected distance of  about 10 kpc north and south  from the central region, while the low velocity components, from about -400 to +300 km s$^{-1}$, were seen as weaker  emission and  at smaller projected distances to the center (see Figure 4).  
On the other hand,  high velocity blue-shifted components were even weaker than the low velocity ones and were  hardly detected in this observation.  
The asymmetry in  spatial and velocity distribution of O$_2$ emission, including the absence of O$_2$ emission along the line of sight around the nuclear region,  can be attributed to the geometry of molecular gas in Mrk 231 and the asymmetry of the molecular outflow from the central AGN, which had been found in CO observations  up to about 3 kpc scale away from the center of Mrk 231 in projection \citep{Feruglio2010, Cicone2012}. 

The absence of a blue-shifted component of   O$_2$ emission may be caused asymmetric distribution of molecular gas in the regions about 10 kpc away from the center, because O$_2$ should be produced by the interaction between the outflowing gas from the AGN center and the local molecular gas there. 
With  velocity of about 500  km s$^{-1}$,  20 Myr are required for gas to travel 10 kpc, which  is the typical projected distance between the  O$_2$  emission regions and the center of Mrk 231.   
Thus O$_2$ emission in Mrk 231 traces the outflow from the central AGN on a characteristic dynamic  time  scale of 20 Myr.   

   X-ray emission  extended up to 25 kpc away from the center of Mrk 231 was found with half-mega second Chandra spectral imaging.
A merger remnant was proposed as the  main contributor to the large-scale X-ray halo of Mrk 231  \citep{2014ApJ...790..116V}. 
However, the central AGN is responsible for emission at smaller ($\le$ 6 kpc) distances.
We suggest that large scale AGN molecular outflow is the most plausible mechanism to produce a shock that drives the enhancement of the O$_2$ abundance but a merger  is another possibility for producing a shock (or multiple shocks) that enhanced the abundance of O$_2$.

 Models of shock chemistry \citep{Chen2014,Neufeld2014,Melnick2015,2019A&A...622A.100G}  suggest that essentially all of the gas phase oxygen will be incorporated into H$_2$O and CO if the shock velocity exceeds 20 km s$^{-1}$.  Thus, a single shock that provides the velocity offset of the redshifted O$_2$ emission of Mrk 231 could not be responsible for dramatically enhancing the abundance of molecular oxygen.  Rather, a number of lower velocity shocks produced in the outflow, having different velocities relative to the general outflow, could collectively  provide the observed large O$_2$ abundance. 
 A high O/CO ratio up to  2$\times10^4$ can be produced in a model of a UV irradiated molecular shock \citep{2019A&A...622A.100G}. Such irradiated  molecular clouds with shocks located  far away from the center of galaxies   can have physical properties similar to those of  the O$_2$-emitting regions  in Mrk 231.  Chemical networks  coupled with dynamic models including radiation field  are necessary for understanding the high O$_2$/CO abundance ratio there. 

For 15K gas under LTE conditions, the 118.75034 GHz O$_2$ transition is the brightest among the 114 O$_2$ lines  listed between 10 GHz and 1000 GHz \citep{Pickett1998}. In total, only 12 transitions make  meaningful contributions, which is consistent with what had been predicted with an LVG model \citep{Liseau2005}. The total O$_2$ flux amounts to 2.9 times the flux at 118 GHz. Similar procedures applied to 17 transitions of CO yielded a CO total flux 5.1 times that of  CO 1-0. Given the comparable fluxes of the O$_2$ $N_J$=$1_1$-$1_0$  transition and the CO 1-0 line  in outer disk region of Mrk 231, the O$_2$ molecule is comparable to CO as an important coolant of such  molecular gas.


Based on CO observations \citep{Cicone2014,2019MNRAS.483.4586F}, massive molecular outflows, which have negative feedback on star formation in galaxies,   are quite common in AGN. The  O$_2$ $N_J$=$1_1$-$1_0$  transition is thus a promising new tool for studying  such outflows on  dynamic time scales of a few 10s of Myr.  AGNs with molecular gas detected via strong CO emission will be good candidates for searching for  O$_2$ emission  in extragalactic sources.

\section{Summary and future prospects}

With  deep observations toward Mrk 231 using the IRAM 30 meter telescope and NOEMA, we detected O$_2$  $N_J$=$1_1$-$1_0$ emission in external galaxy for the first time, at about the 12 $\sigma$ level. The detected O$_2$ emission is located in  regions about 10 kpc away from the center of  Mrk 231 and  may be caused by the interaction between the AGN-driven molecular outflow and the outer disk molecular clouds. 

Massive molecular outflows are quite common in AGN based on CO observations \citep{Cicone2014}. The  O$_2$ $N_J$=$1_1$-$1_0$  transition could prove to be a new tool for studying the effect of  such outflows, because outflows identified with CO observations can be severely contaminated by emission components of the host galaxy. On the other hand, O$_2$ emission from dense molecular gas in host galaxy can be neglected, which means  O$_2$ emission can be used to study a molecular outflow even if it is perpendicular to the line of sight. AGNs with molecular outflows detected in CO emission are good candidates for detecting O$_2$ emission in extragalactic sources. Even though Mrk 231, a northern source at Declination of $\sim$ 57$^{\circ}$, cannot be reached with ALMA, other red-shifted AGNs with molecular outflows may be studied with high resolution high sensitivity O$_2$  $N_J$=$1_1$-$1_0$ observations using ALMA and NOEMA, as well as the  Next-Generation  Very Large Array (ngVLA) in the next decade. O$_2$  lines at submillimeter  wavelengths,  such as that at 424.76 GHz, with suitable red-shift shifting it to  an observable window of ALMA, may be useful for studying molecular oxygen in more distant galaxies. 

\acknowledgments

This work is supported by the National Natural Science Foundation of China grant 11988101, 11590783,  11590782, 11725313, and 11690024,  and  the International Partnership Program of Chinese Academy of Sciences grant No. 114A11KYSB20160008. This study is based on observations carried out under project number 068-15 with the IRAM 30m telescope and d17ae001 with the NOEMA. IRAM is supported by INSU/CNRS (France), MPG (Germany) and IGN (Spain). This research was carried out in part at the Jet Propulsion Laboratory which is operated for NASA by the California Institute of Technology.  YG's research is supported by National Key Basic Research and Development Program of China (grant No. 2017YFA0402704), National Natural Science Foundation of China (grant Nos. 11861131007, 11420101002), and Chinese Academy of Sciences Key Research Program of Frontier Sciences (grant No. QYZDJSSW-SLH008). This work also benefited from the International Space Science
Institute (ISSI/ISSI-BJ) in Bern and Beijing, thanks to the funding of
the team ?Chemical abundances in the ISM: the litmus test of stellar
IMF variations in galaxies across cosmic time? (Principal Investigator
D.R. and Z-Y.Z.)     We thank the anonymous referee for helpful comments, which improved the manuscript.

\begin{figure*}
\centerline{
\includegraphics[width=0.8\textwidth]{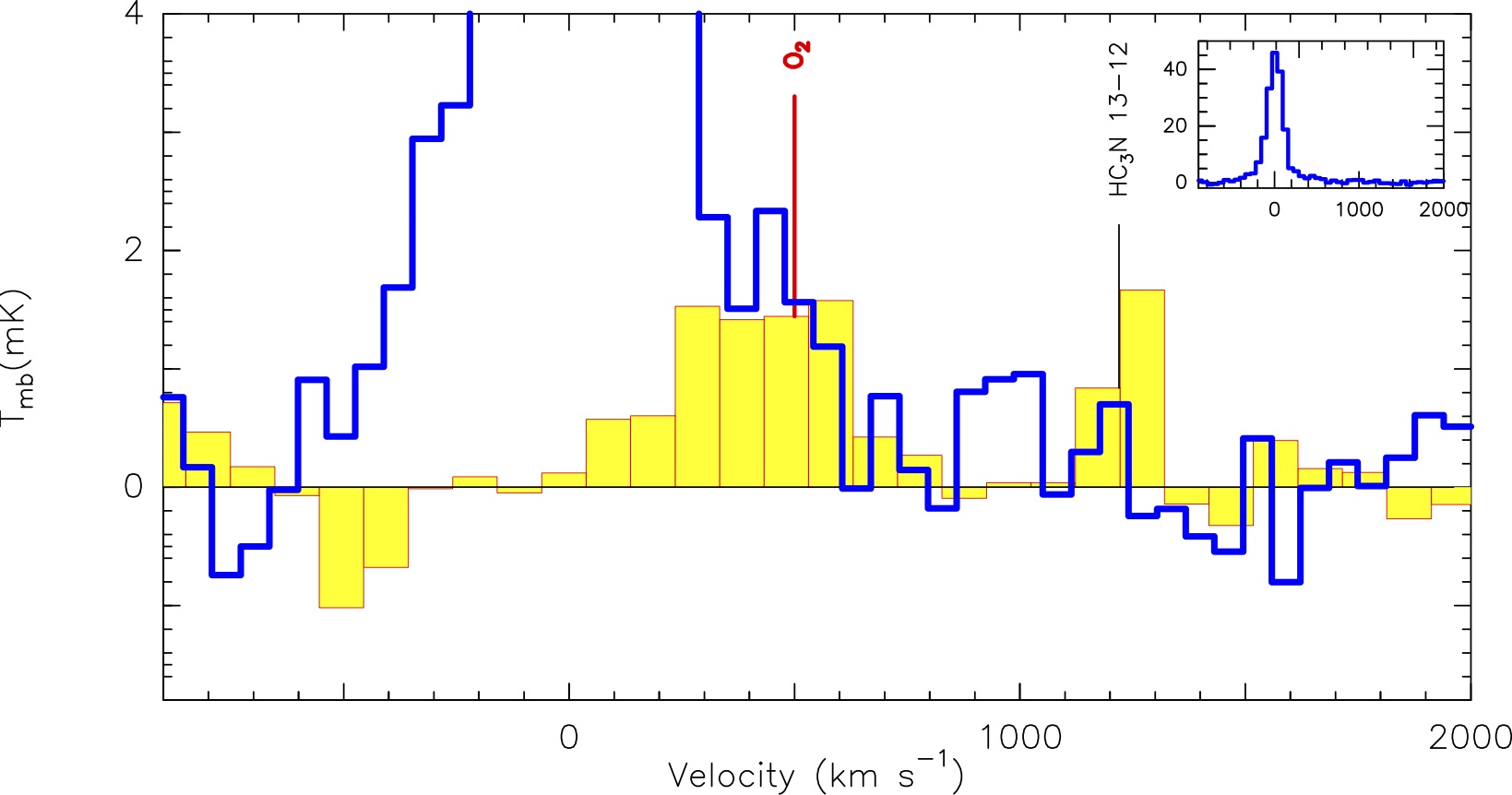}
}
\vspace{-1mm}
\caption{\textbf{O$_2$ $N_J$=$1_1$-$1_0$ line  emission having   rest frequency   118.75034 GHz  (filled yellow histogram and red line) in Mrk 231 observed with the IRAM 30 meter telescope. The spectrum combins all data obtained in 4 days. The  $x$  axis is   radio-defined velocity, which is referred to the frequency of red-shifted O$_2$  as   118.75034/(1$+z$)GHz with $z$=0.04217.  The velocity resolution presented here has been smoothed to 98.66 km s$^{-1}$.   The blue line is the  CO 1-0 data obtained simultaneously, aligned  with radio-defined velocity refereed to the frequency of red-shifted CO 1-0  as   115.271202/(1$+z$)GHz.  The HC$_3$N $J$=13-12 line having rest frequency of 118.2707 GHz was observed with and is plotted on the same velocity scale as the O$_2$.  It appears at a velocity $\simeq$ 1200 km s$^{-1}$ as the emission velocity of this dense gas tracer is close to 0 km s$^{-1}$, the systemic velocity of Mrk 321.  The  $y$  axis is the main beam temperature in milli-Kelvin  for both O$_2$  and CO 1-0 spectra. The spectrum in the insert box (blue line) gives a  view of the complete CO line over the same velocity range as in the main figure.}}
\end{figure*}

\begin{figure*}
\centerline{
\includegraphics[width=0.65\textwidth]{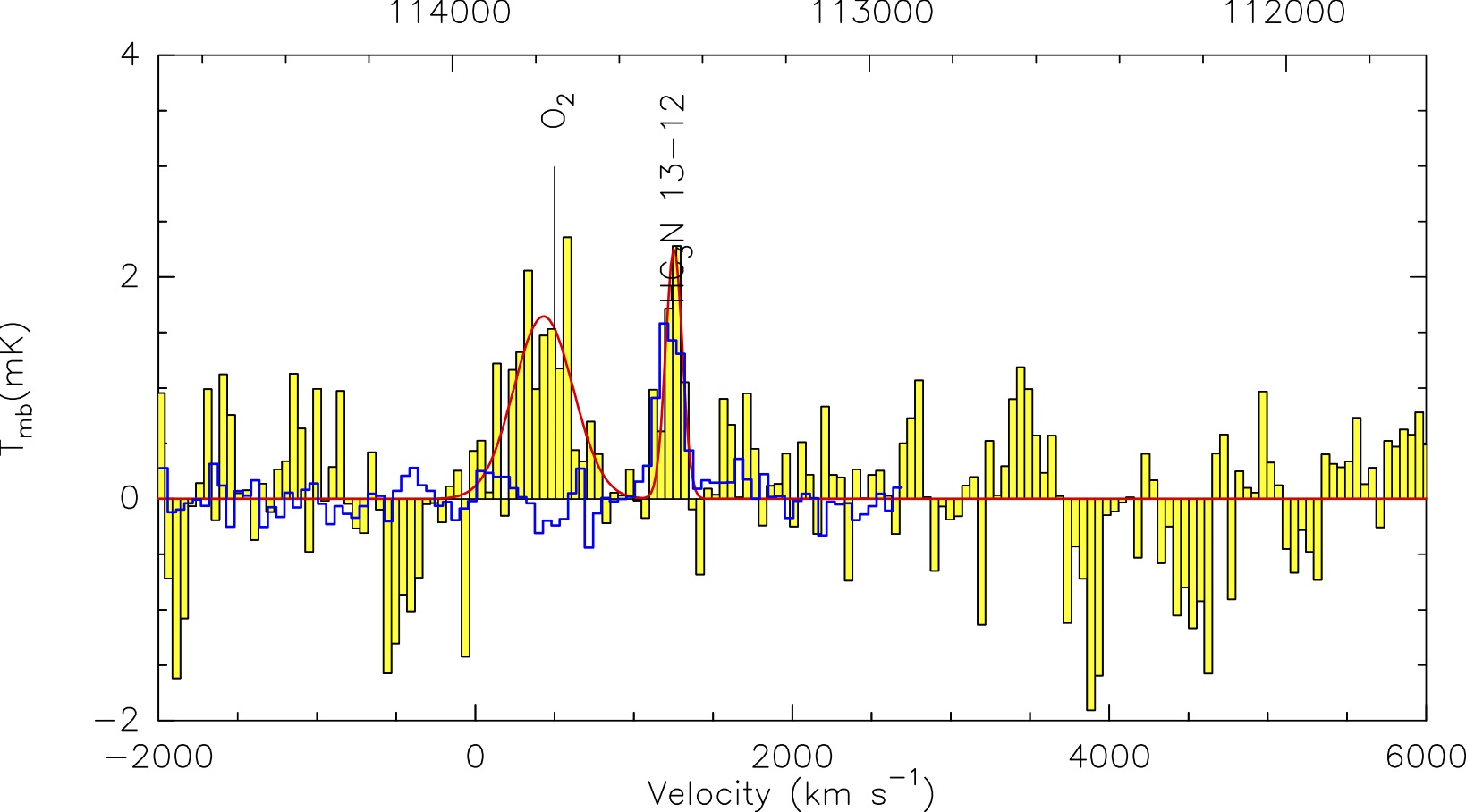}
}
\vspace{1mm}
\caption{ \textbf{Spectrum (black and filled yellow)  in Mrk 231 with the same data shown in Figure 1 presented with less velocity smoothing and broader velocity coverage. The $x$ axis is velocity  as defined in Figure 1. The  rms of this spectrum is  0.65 mK at a frequency resolution of 18.75MHz, corresponding to velocity resolution 49.33 km s$^{-1}$.  The red line is the result of two-component Gaussian fitting. The velocity coverage is from -2000 to 6000 km s$^{-1}$, which represents about 3000 MHz frequency coverage as indicated by the frequency scale (in MHz) at the top of the figure. The overlaid blue spectrum is from the central region of Mrk231 obtained with NOEMA, which is the same as that in the upper left in Figure 4.}}
\end{figure*}

\begin{figure*}
\centerline{
\includegraphics[width=0.45\textwidth]{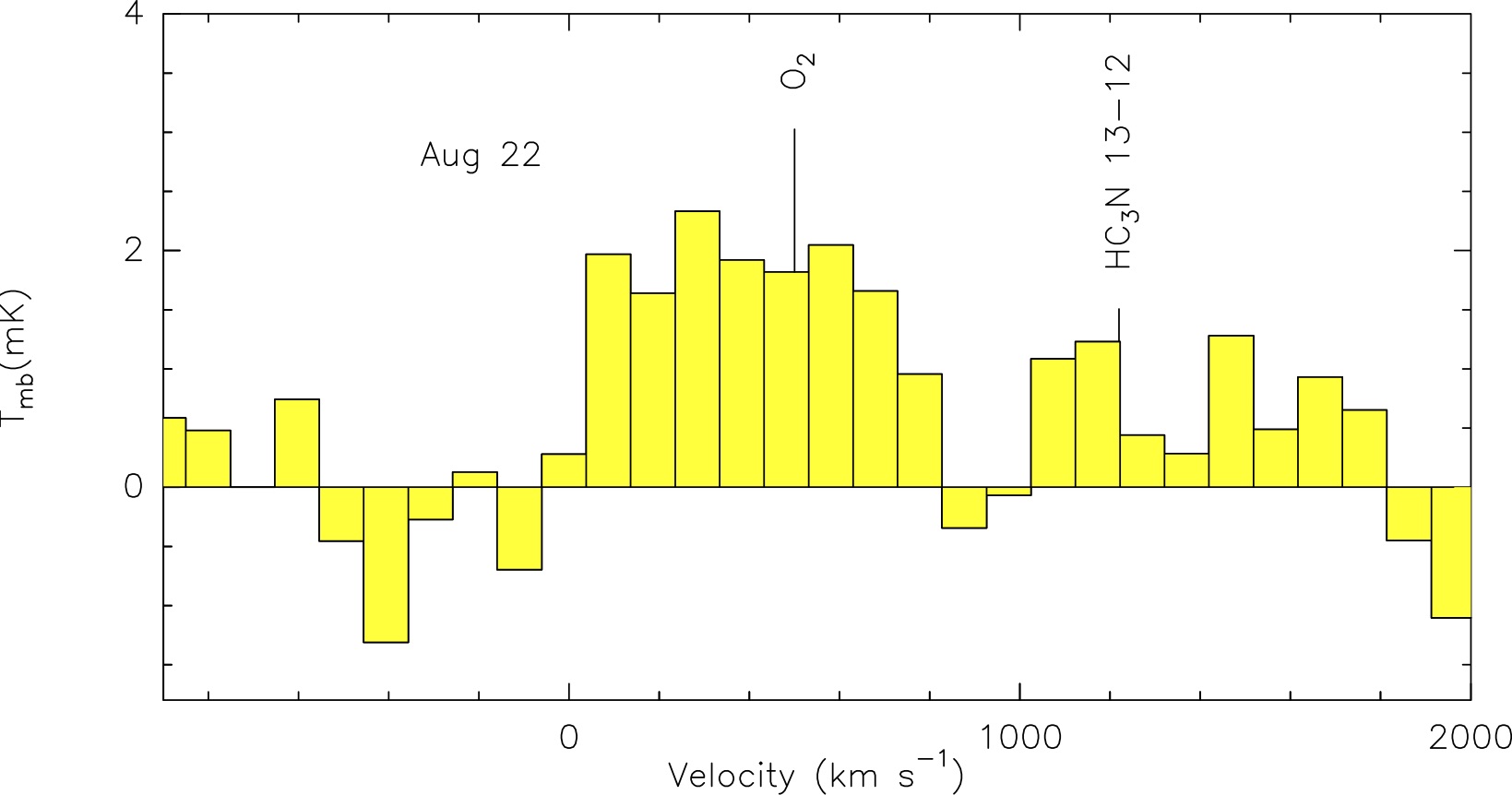}
\includegraphics[width=0.45\textwidth]{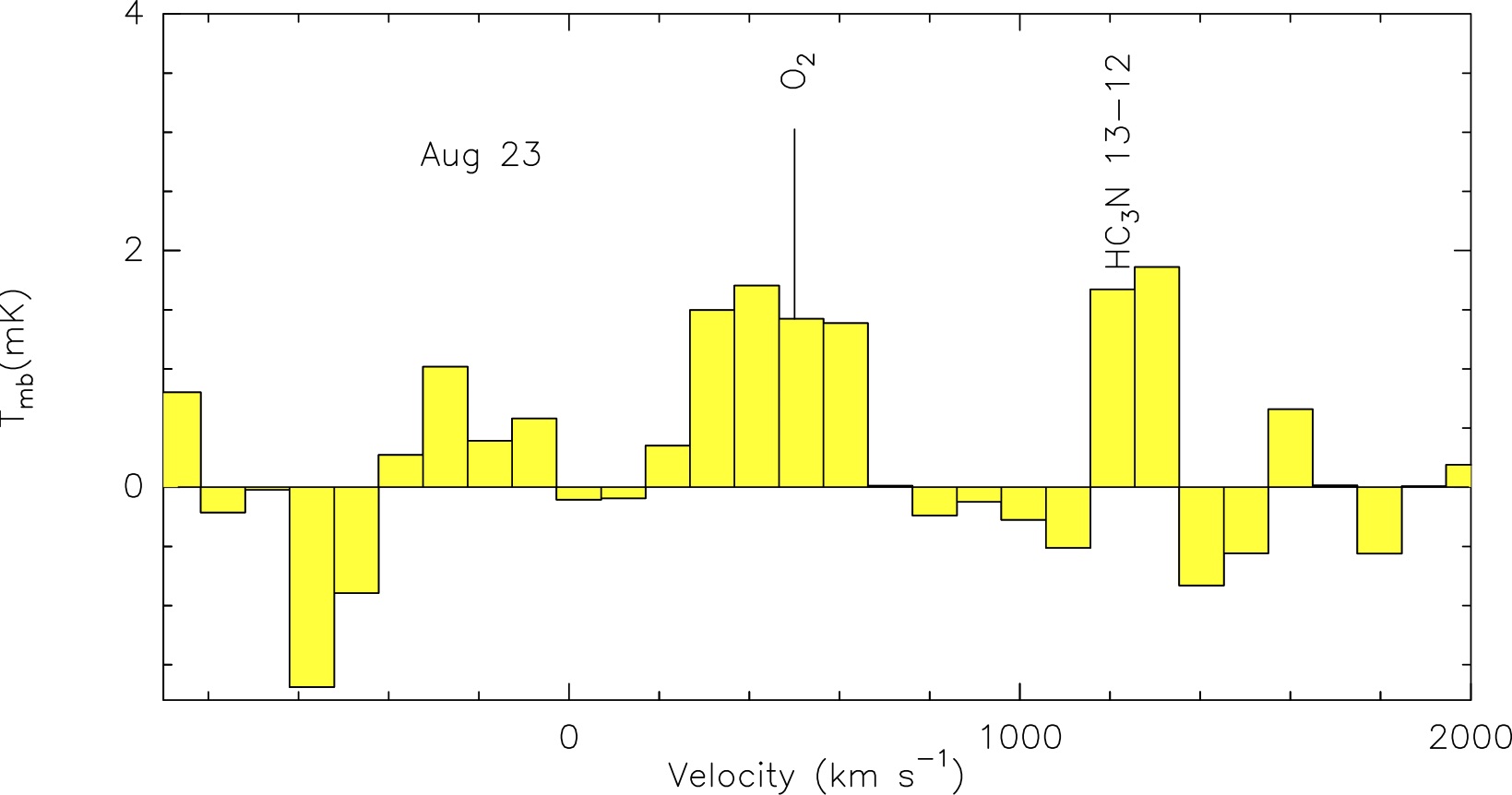}
}
\vspace{1mm}
\caption{\textbf{Spectra in Mrk 231 observed with the IRAM 30m telescope  on Aug 22 (left), and Aug 23 (right), 2015.}}

\end{figure*}

\begin{figure*}
\centerline{
\includegraphics[width=0.8\textwidth]{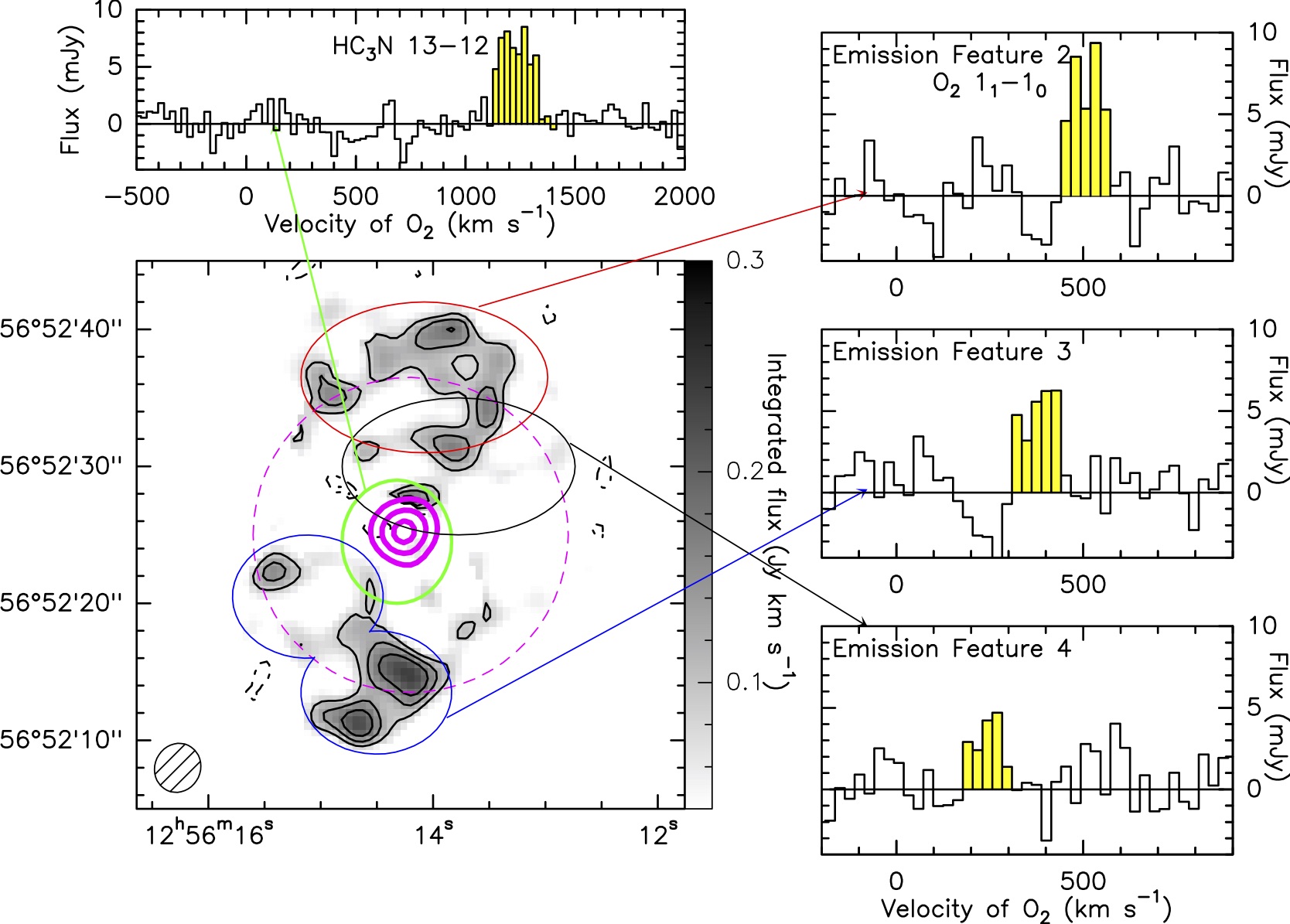}
}
\vspace{-0mm}
\caption{\textbf{O$_2$ emission detected with the IRAM NOEMA interferometer. The grey scale and contours in black  are the velocity integrated  O$_2$  emission in  units of Jy km s$^{-1}$, integrated over 180 to 570 km s$^{-1}$. The  grayscale colorbar at the right is in units of  Jy km s$^{-1}$.  The contours are from 2$\sigma$ increasing in steps of  1$\sigma$, which  corresponds  to 0.045Jy km s$^{-1}$. The contours in magenta are the velocity-integrated  HC$_3$N 13-12 emission starting from  and with  steps of  0.4 Jy km s$^{-1}$. The ellipses are the regions over which data were integrated to obtain the  spectra, including that for  HC$_3$N 13-12 emission,  with  the same definition  of  velocity as  in Figure 1. The velocity resolution is 26.3 km s$^{-1}$ for the spectra of O$_2$  and HC$_3$N 13-12 presented here. The dashed circle is the main beam size of IRAM 30 meter and the filled grey ellipse at bottom left is the beam size of  NOEMA data (3.63$''\times$3.32$''$, PA=-65$^{\circ}$).}}

\end{figure*}

\begin{figure*}
\centerline{
\includegraphics[width=0.3\textwidth]{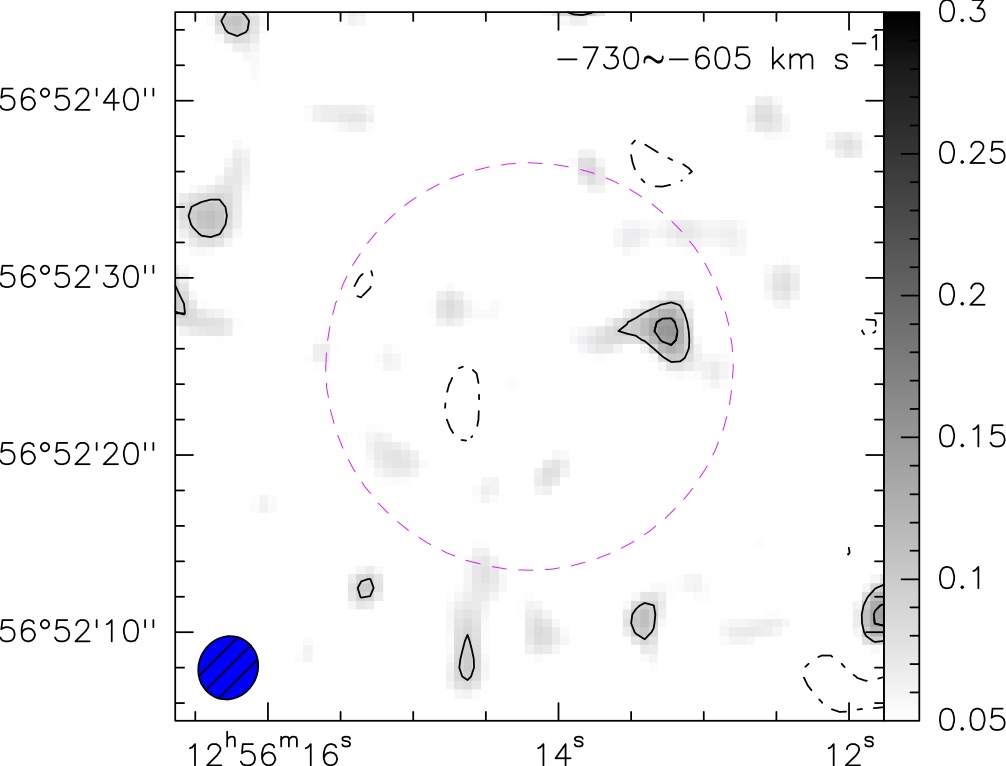}
\includegraphics[width=0.3\textwidth]{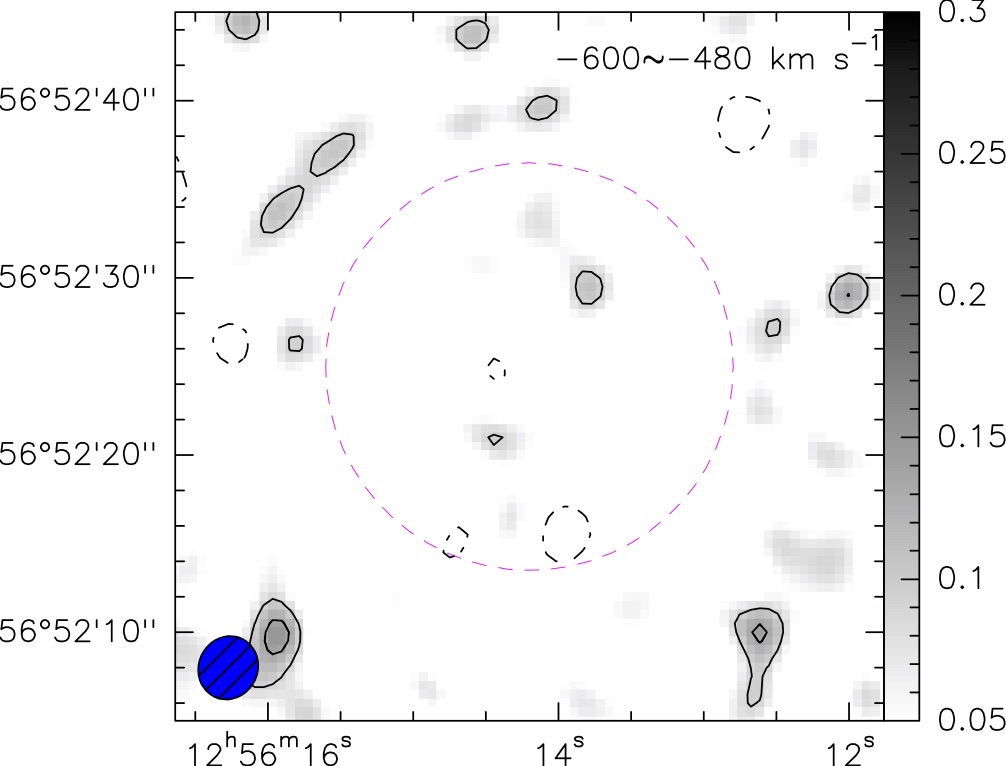}
\includegraphics[width=0.3\textwidth]{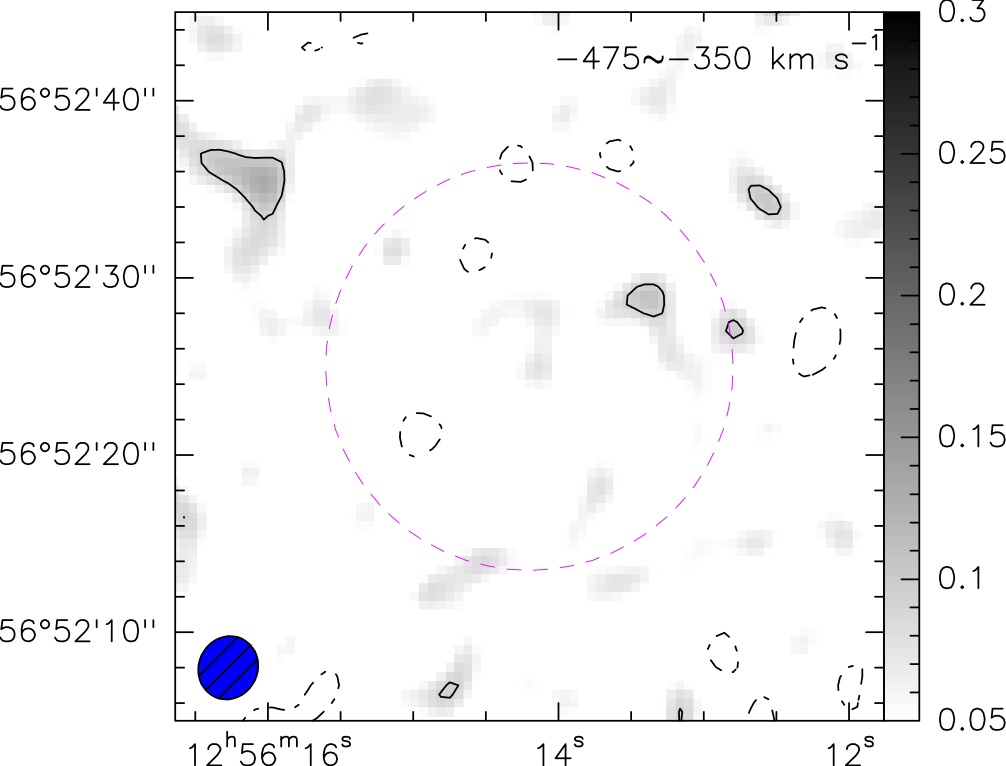}
}
\centerline{
\includegraphics[width=0.3\textwidth]{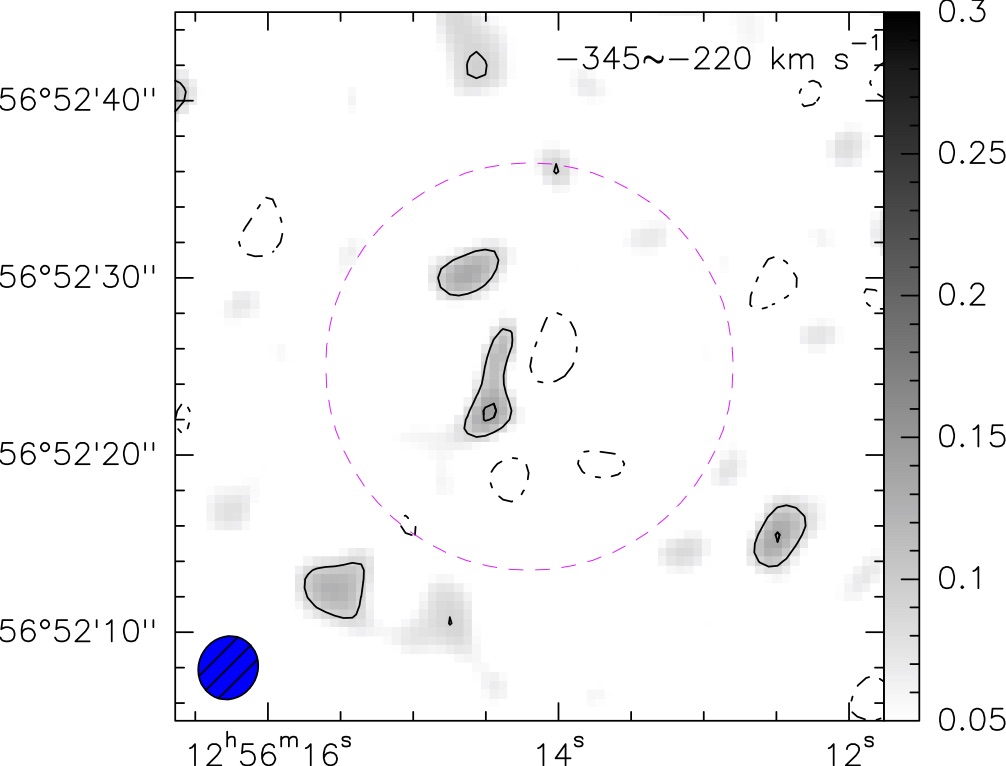}
\includegraphics[width=0.3\textwidth]{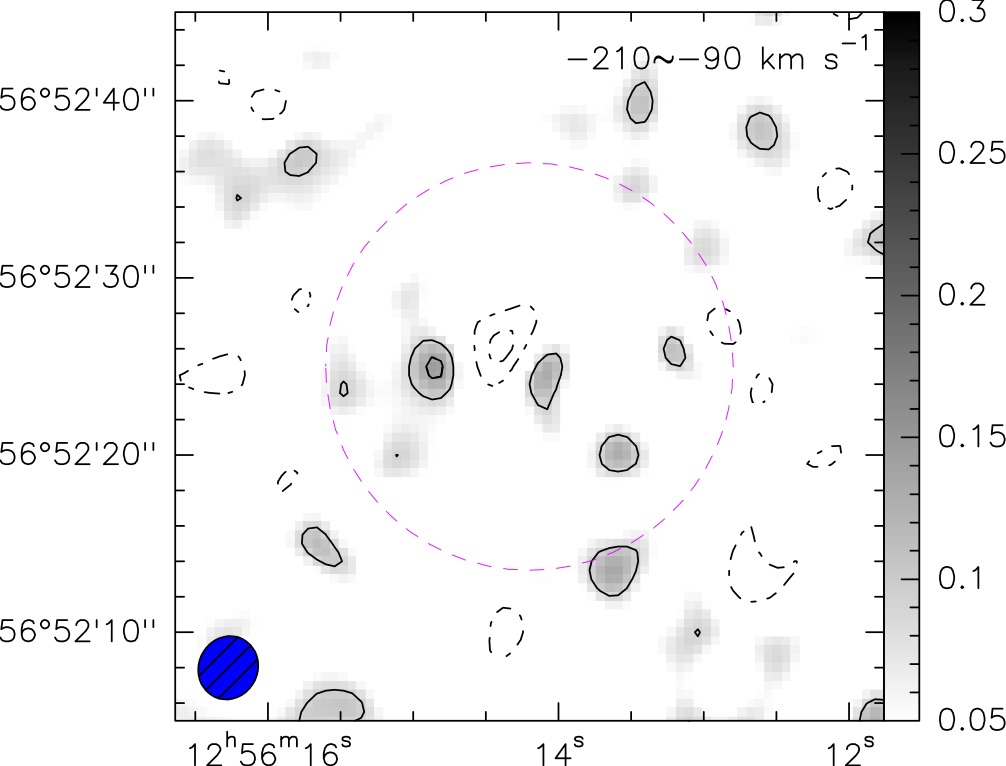}
\includegraphics[width=0.3\textwidth]{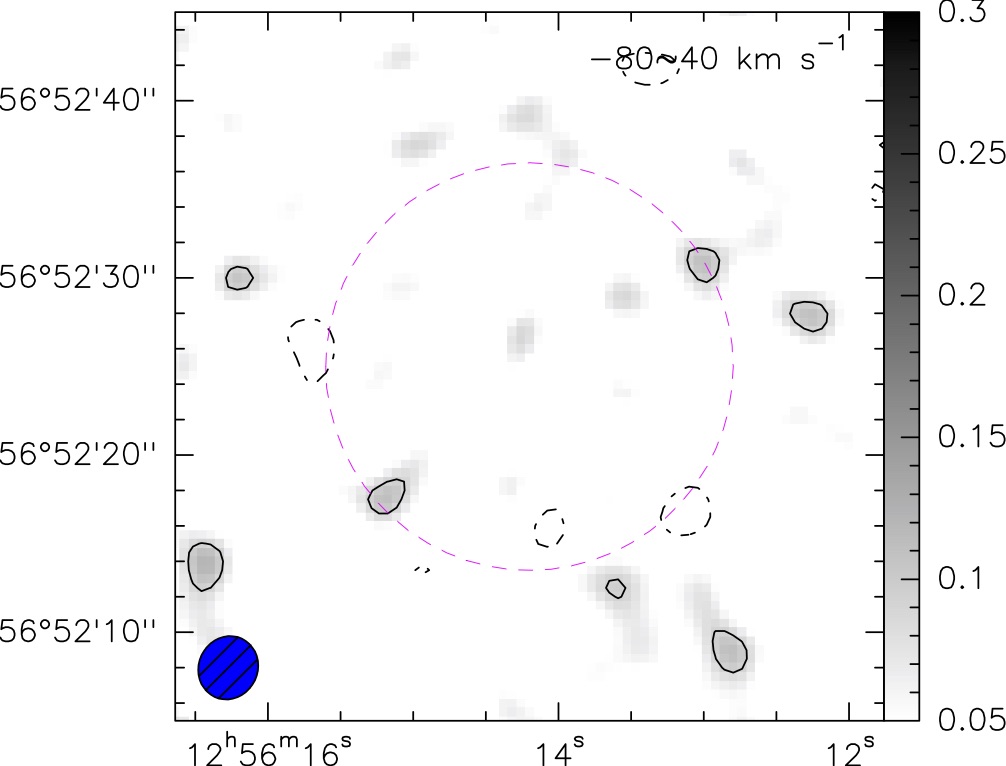}
}
\centerline{
\includegraphics[width=0.3\textwidth]{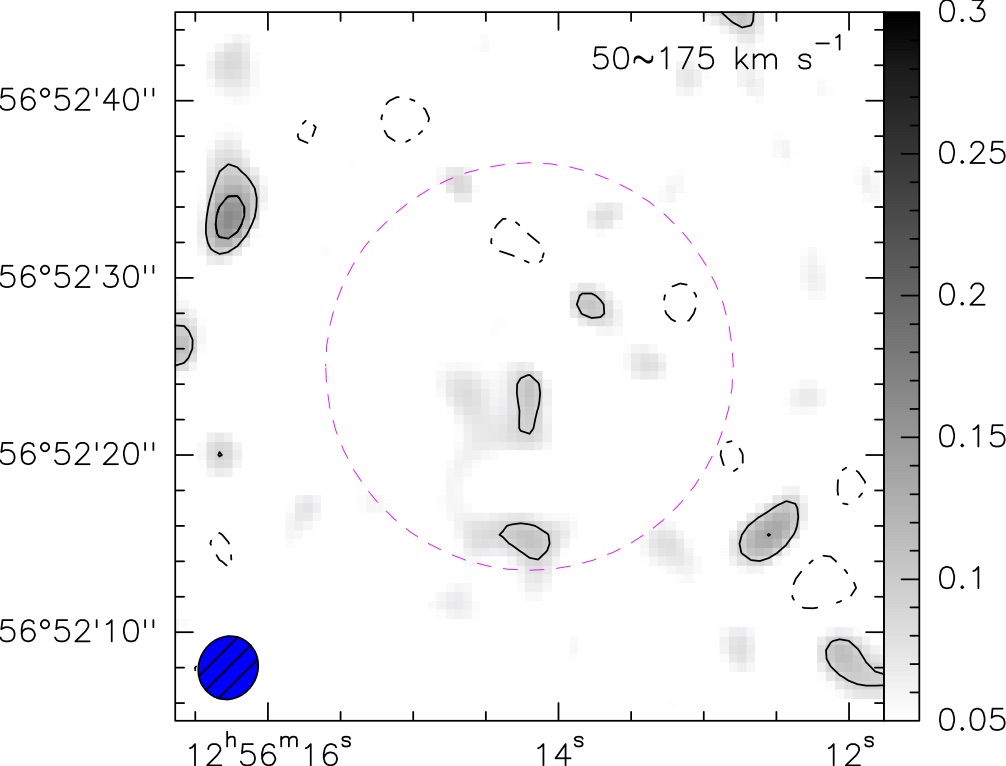}
\includegraphics[width=0.3\textwidth]{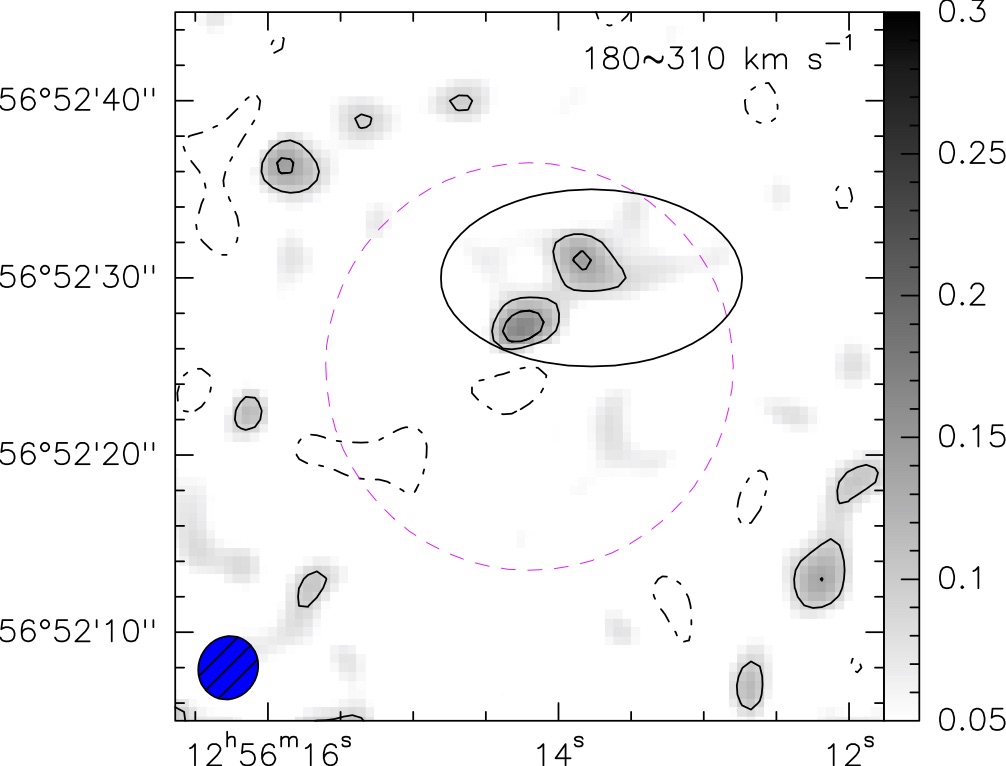}
\includegraphics[width=0.3\textwidth]{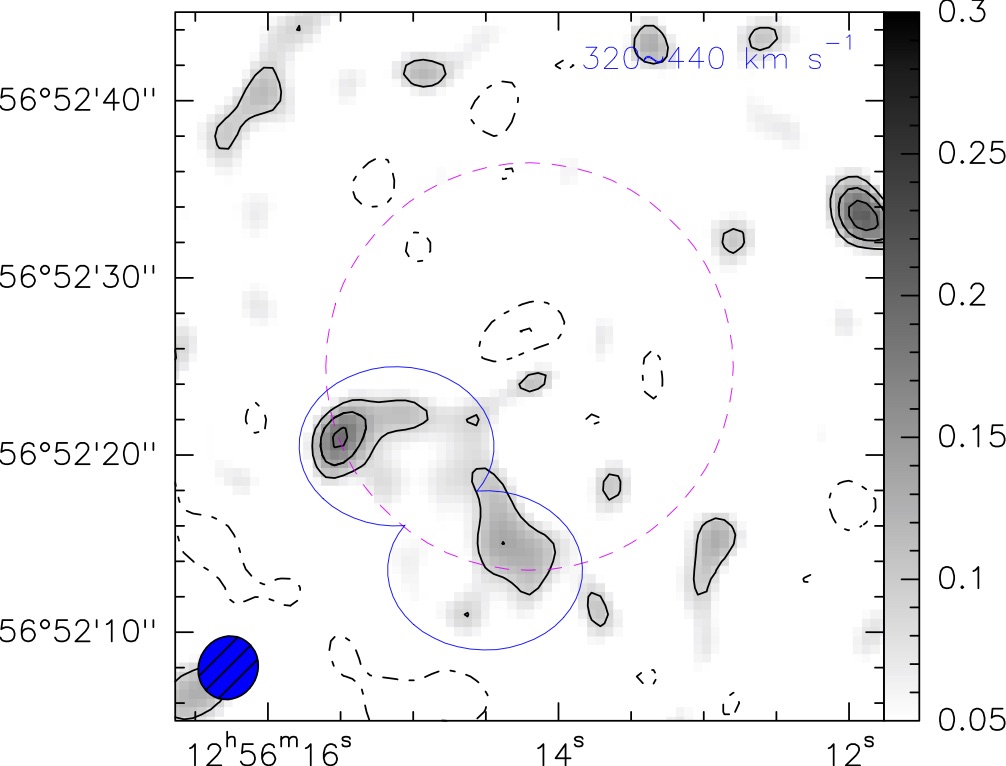}
}
\centerline{
\includegraphics[width=0.3\textwidth]{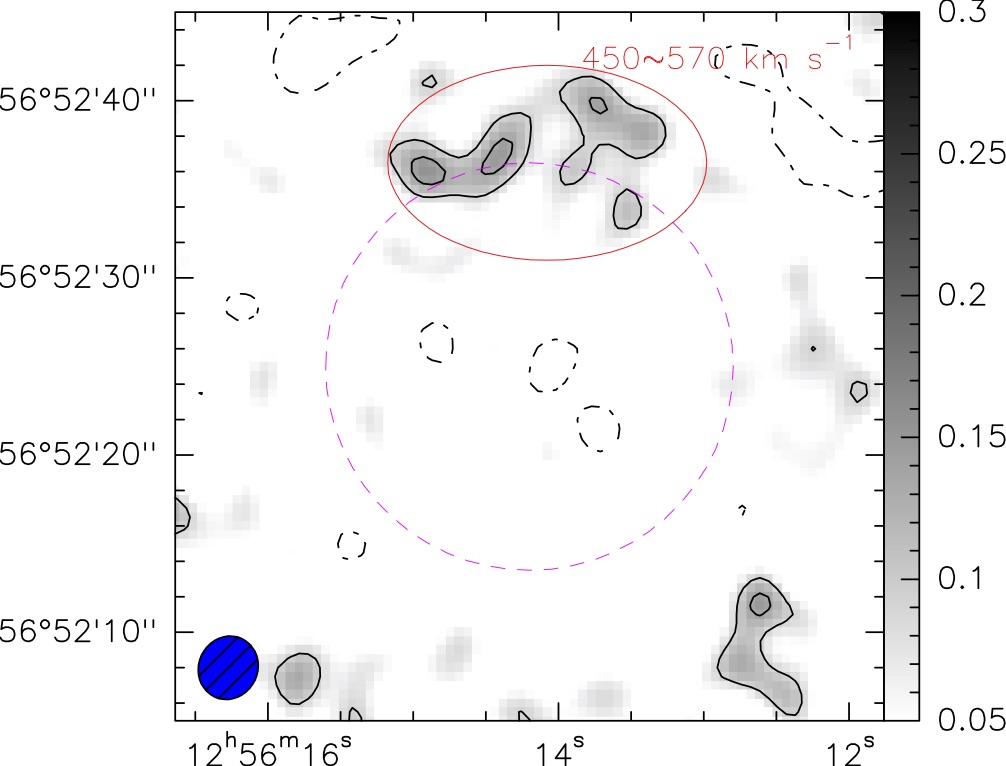}
\includegraphics[width=0.3\textwidth]{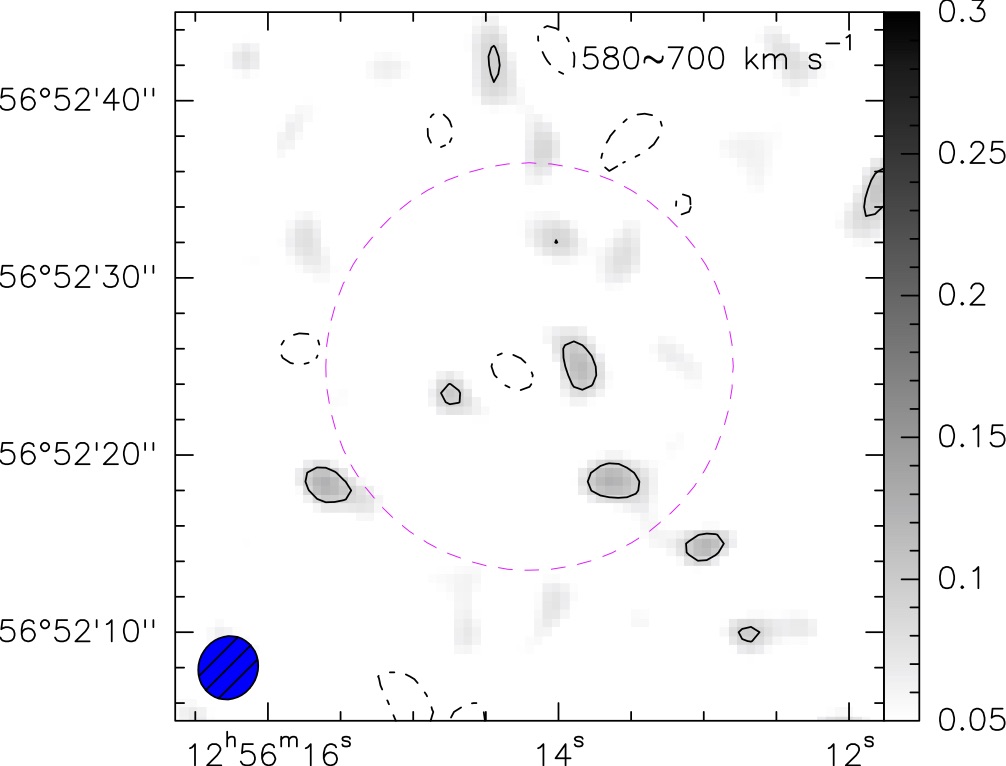}
\includegraphics[width=0.3\textwidth]{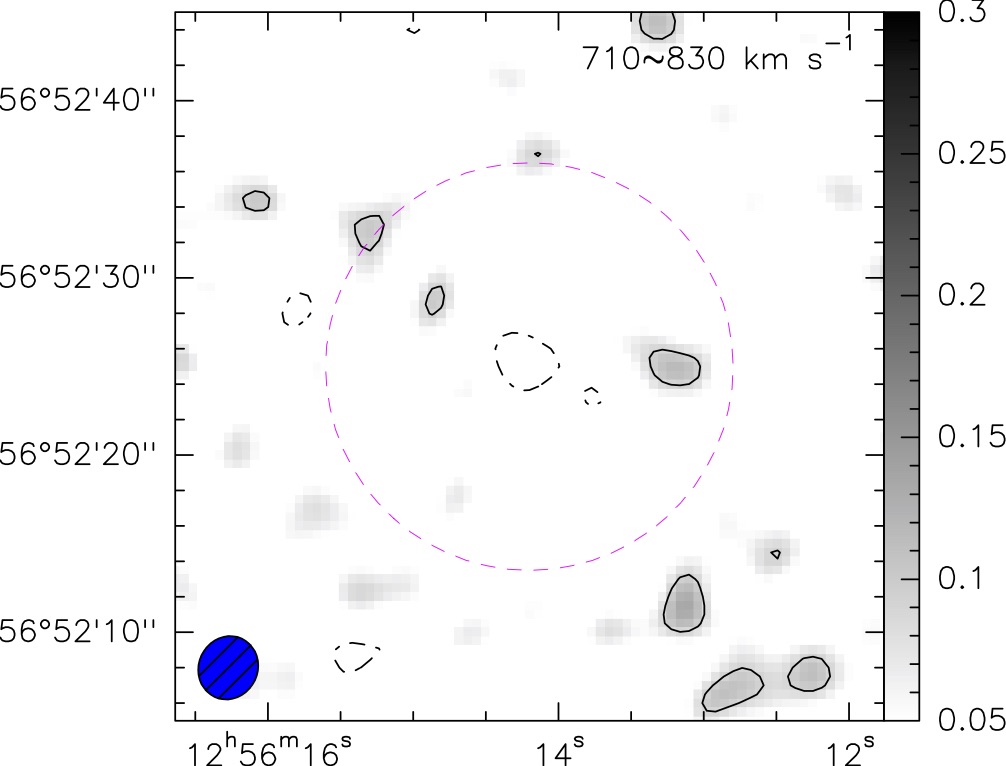}
}
\vspace{2mm}
\caption{ \textbf{Velocity integrated maps of O$_2$ emission for different velocity ranges with the IRAM NOEMA data.  The contours are from 2$\sigma$ increasing in steps of  1$\sigma$, which  corresponds  to 0.045 Jy km s$^{-1}$.   The color-bars, circle and ellipses   are the same as  in Figure 4.}}

\end{figure*}

\begin{table*}[b]
\begin{center}{ \bf Table 1. Emission features of O$_2$ $N_J$=$1_1$-$1_0$ detected with the IRAM 30 meter and NOEMA }\end{center}
\small
\begin{center}
\begin{tabular}{clllllllllllll}
\hline
 Emission Feature&Flux    & Line Center &Line Width &Telecope  \\
          &    Jy km s$^{-1}$       &    km s$^{-1}$            &   km s$^{-1}$  \\
\hline                                                            
1  & 3.7$\pm$0.3  &418.7$\pm$47.9 & 456.0$\pm$107.1& 30 meter\\
2   & 0.80$\pm$0.13 &  511.5$\pm$8.1&   93.5$\pm$13.1&  NOEMA\\
3   & 0.70$\pm$0.12 &   391.5$\pm$9.7 &    98.8$\pm$16.7&  NOEMA\\
4   & 0.36$\pm$0.10 &   248.5$\pm$11.9 &    77.6$\pm$24.7 & NOEMA\\
\hline
\end{tabular}
\end{center}
Note: The errors only include the deviation  from Gaussian fitting. About 10 to 20\% of the velocity-integrated flux should be also considered as the uncertainty for comparison with other observations, arising from the uncertainties in the absolute flux calibration and pointing errors for the single dish observations.  Emission features 2, 3, and 4 correspond to the three spectra in the right of Figure 4, while emission feature 1 corresponds to the spectrum in Figures 1 and 2. 

\end{table*}

\begin{table*}[b]
\begin{center}{ \bf Table 2.  O$_2$ detections and best upper limits}\end{center}
\small
\begin{center}
\begin{tabular}{lcrlllllllllll}
\hline
Source &  O$_2$ to H$_2$ abundance ratio &Telescope&  References\\
\hline                                                            
Orion &1$\times10^{-6}$ &Herschel &\cite{Goldsmith2011}\\
 $\rho$ Oph &5$\times10^{-8}$ &Odin&\cite{Larsson2007}\\
 NGC 6240 &$<1\times10^{-6}$ &IRAM 30 meter & \cite{Combes1991}\\
 B0218+357 &$<2\times 10^{-7}$ &NRAO12 meter & \cite{Combes1997}\\
 Mrk231 (central 2kpc) &$<8\times10^{-8}$& IRAM NOEMA& This work\\
 Mrk231 (outflow) &$>1\times10^{-4}$& IRAM 30 meter and NOEMA&This work\\

\hline
\end{tabular}
\end{center}
Note: We present the CO 1-0 to O$_2$ line ratio to estimate  O$_2$/H$_2$ abundance ratio only for sources with detected CO emission or an upper limit (Mrk 231 O$_2$ emitting region). 1$\sigma$ is used for  upper or lower limits. The O$_2$/H$_2$ abundance ratios in the  two Galactic sources (Orion and $\rho$ Oph) were estimated using optically thin isotopologues instead of CO, while  the ratio for B0218+357 was estimated based on an absorption spectrum. 
\end{table*}

\end{document}